# Stapledon's Interplanetary Man: A Commonwealth of Worlds and the Ultimate Purpose of Space Colonisation

I. A. Crawford
Department of Earth and Planetary Sciences, Birkbeck College, Malet Street, London, WC1E 7HX, UK

> *"The stars give no answer. But within ourselves, and in one another, and in our unity together, the answer lies ...."* [1; p. 59]

**Abstract**

In his 1948 lecture to the British Interplanetary Society Stapledon considered the ultimate purpose of colonising other worlds. Having examined the possible motivations arising from improved scientific knowledge and access to extraterrestrial raw materials, he concludes that the ultimate benefits of space colonisation will be the increased opportunities for developing human (and post-human) diversity, intellectual and aesthetic potential and, especially, 'spirituality'. By the latter concept he meant a striving for "sensitive and intelligent awareness of things in the universe (including persons), and of the universe as a whole." A key insight articulated by Stapledon in this lecture was that this should be the aspiration of all human development anyway, with or without space colonisation, but that the latter would greatly increase the *scope* for such developments. Another key aspect of his vision was the development of a diverse, but connected, 'Commonwealth of Worlds' extending throughout the Solar System, and eventually beyond, within which human potential would be maximised. In this paper I analyse Stapledon's vision of space colonisation, and will conclude that his overall conclusions remain sound. However, I will also argue that he was overly utopian in believing that human social and political unity are prerequisites for space exploration (while agreeing that they are desirable objectives in their own right), and that he unnecessarily downplayed the more prosaic scientific and economic motivations which are likely to be key drivers for space exploration (if not colonisation) in the shorter term. Finally, I draw attention to some recent developments in international space policy which, although probably not influenced by Stapledon's work, are nevertheless congruent with his overarching philosophy as outlined in 'Interplanetary Man?'

**1. Introduction**

In his 1948 lecture on "Interplanetary Man?" Stapledon [2] was primarily concerned with the socio-political underpinning, and the ultimate purpose, of space exploration and colonisation. Note the question mark in the title – Stapledon appears not to have believed that that a human expansion into the Universe is inevitable, but is likely to occur only if certain physical and/or societal conditions are met. As one would expect from a thinker of Stapledon's stature, his lecture was rich in philosophical content, not all of which may have immediately appealed to his audience of early space enthusiasts, and not all of which may appeal to 'rocket scientists' today. Nevertheless, while accepting that Stapledon unnecessarily downplayed the more prosaic scientific and economic motivations for space exploration, I will argue that his lecture contained real insights into the wider socio-cultural context of space activity that need to be considered by anyone with a serious interest in the future of humanity in space.

In what follows I first attempt to summarise the content of Stapledon's 1948 lecture, using the sub-divisions in the published version [2] as a guide, and offer some personal commentary of my own. I will conclude by summarising what I consider to have been Stapledon's key insights as expressed in his lecture, and briefly discuss the relevance of his thought for the development of a 21st Century space policy.

All quotations are from "Interplanetary Man?" unless otherwise stated.

**2. Socio-political foundations**

In his Introduction, Stapledon notes that the pace of human cultural and technological development is accelerating at an ever increasing rate and that the future, and probably the near future, will be very different from the past, even though we cannot see what this future will be like. As he puts it:

> *"The river of human life has reached a precipice. The cataract plunges – whither?"*

However, he goes on to identify what he sees as three possible near-term futures for humanity:

- Speedy (self-inflicted) annihilation

- Creation of a world-wide tyranny (and implied stagnation)

- The founding of a "new kind of human world" where "with a modicum of wisdom" everyone has "the chance to develop and express such capacity as he has for truly human living and truly human work in the great common enterprise of man."

However, this list of alternative futures is somewhat problematic for a number of reasons. The first bullet point probably seemed all too likely in 1948 and was, and indeed still is, a possible near-term end point for human technological civilisation. That said, the actual self-extinction of *Homo sapiens* as a species will be a lot harder to achieve, and may be all-but impossible (as Stapledon himself recognized at the end of Chapter V of *Last and First Men* [3; p.100]), so this worst-case scenario is unlikely to be the (near term) fate of humanity.

The second possible future probably also seemed a real danger in 1948. Again, it is not a desirable outcome, but I think Stapledon was wrong to imply that tyranny *per se* is an endpoint of human history. For one thing, it seems inconceivable that any tyranny, no matter how oppressive, could last forever, or indeed that its duration could be anything other than a very small fraction of humanity's total existence. Moreover, there are gradations of possible tyrannies, and not all may be incompatible with technological development or, in the present context, space exploration. Indeed, Cockell [4] has drawn attention to the unfortunate fact that the space environment may be especially conducive to generating tyrannies, and unless we are very careful the expansion of humanity into the universe may occur under totalitarian socio-political systems. While clearly not desirable, such an eventuality would not, as Stapledon appears to imply, be incompatible with the human colonisation of space.

The third of these possible futures is clearly the most desirable, but it is in a different category from the others. Whereas annihilation, by definition, would be the end of the story for humanity, and global tyranny would be a (non-desirable but presumably transient and evolvable) political system, Stapledon's third option is really a utopian socio-political *aspiration* that would be compatible with a variety of social and political arrangements. Stapledon's own preferred political arrangements for "a new kind of human world" are not explicitly specified in '*Interplanetary Man?*' However, as we shall see, some kind of world government is implied, and in other work (e.g. *Last and First Men* [3] and *Old Man in New World* [1]) Stapledon was clearly sympathetic to some form of world government as an appropriate political structure for a united humanity. As he was also opposed to 'global tyranny' it follows that he believed that a non-tyrannous (i.e. democratic and presumably federal) world government is both possible and desirable.

Stapledon goes on to imply that space colonisation will only occur, and perhaps *should* only occur, within the latter socio-political context. However, while everyone will agree that of the three possible futures outlined by Stapledon this is the most attractive, subsequent history has shown that socio-political utopias are not a prerequisite for space exploration. Indeed, for better or worse, much of the history of space exploration to-date has been driven by nationalistic competition within the context of a politically divided world in which there is very little evidence for the "modicum of wisdom" that Stapledon rightly thought to be desirable. Actually, Stapledon himself acknowledged that at least the early years of space exploration might proceed on a nationalistic basis when he remarked:

> "*Alas! Must the first flag to be planted beyond Earth's confines be the Stars and Stripes, and not the banner of a united Humanity?*"

That said, I think Stapledon was right to point to synergies between human political unification and space exploration, especially in the longer term. Indeed, the link between the two had already been made in an earlier talk to the Society (October 1947) by A. V. Cleaver entitled "The Interplanetary Project" [5], and it would be interesting to know if Stapledon was aware of Cleaver's earlier contribution. My own thoughts on the relationship between world government and space exploration are set out in detail elsewhere [6], where I suggested that a symbiotic relationship may develop between the two. Specifically, I argued that a united world would have more resources available for space exploration (in part because of the reduced requirements for military budgets), but that space exploration would increasingly provide a 'cosmic perspective' on human affairs which would reinforce the sense of humanity as a single species occupying a small planet and thereby enhance the perceived legitimacy of global government. As I put it then [6]: "a world government may find an ambitious space programme to be desirable for social reasons, but, equally, only a world government will be able to organise one on the necessary scale."

Regardless of whether these speculations turn out to be right or wrong, in the present context we are concerned with Stapledon's thinking and there is no doubt that *he* viewed human unification as an important prerequisite for the colonisation of other planets, as he goes on to ask:

> *"Suppose that mankind has at last become effectively united, both politically and socially. Then what should a united mankind do with the planets?"*

He realised that the answer will depend on the physical state of the other planets in the Solar System, and especially on whether or not they are already inhabited by indigenous intelligent species. As he put it:

> *"Much depends on the conditions of the planets that he visits …. Either man will find elsewhere in the solar system other intelligences, or he will not."*

## 3. If the planets are inhabited

In the case that humanity finds other intelligent species occupying other planets in the Solar System, Stapledon identifies only two possibilities – conflict between humanity and these intelligent races or cooperation with them. In principle there might be other possibilities, for example a kind of 'cold war' stand-off based on fear and/or mutual incomprehension. Stapledon does not discuss alternatives of this kind, but in truth they would unlikely to be stable over the long term so his analysis is probably basically sound.

Stapledon argues that conflict would probably be inevitable unless humanity has managed to unify itself before contact is made (and this is another implicit recognition that, despite the desirability of such unification, it might not occur before space exploration has begun). He even argues that interplanetary conflict might have a short-term beneficial effect on humanity by acting as an additional stimulus for unification. However, he argues that ultimately there can only be two possible outcomes of such a 'War of the Worlds': either the extinction of humanity or the extinction of the other intelligent species, and that neither is desirable.

On the other hand, if humanity can get its own house in order, there are other possibilities:

> *"If ... man does soon succeed in unifying his world society, then it is at least conceivable that some kind of mutually profitable symbiosis with intelligent races on other planets might be established."*

Interestingly, this implies that the other intelligent races would themselves already be politically unified, for otherwise the united humanity would have to interact with disunited aliens, with renewed potential for confusion and conflict. The fact that Stapledon neglects this point indicates the extent to which he implicitly assumed that planets inhabited by intelligent species would be politically united because this is the only rational way of organising planetary affairs (either for humans or non-humans). Clearly, cooperation would be the most desirable outcome, and would enable a start to be made on developing the 'Commonwealth of Worlds' that he develops later in his lecture (see Section 8 below). Unfortunately (or perhaps fortunately, given that humanity is not yet politically unified), by 1948 it was already becoming clear that the Solar System is most unlikely to be inhabited by alien races of comparable intelligence to *Homo sapiens*, and we can now be certain that it is not.

Nevertheless, Stapledon's discussion on the interaction between humanity and other alien races remains relevant in a wider, galactic context. For the same basic perspective applies:

either man will find elsewhere in the Galaxy other intelligences, or he will not. As Stapledon himself explores in *Starmaker* [7], interactions between intelligent races in the Galaxy may result in conflict or cooperation, and probably both at different times and places. Indeed, it is the lack of evidence for interstellar colonisation and conflict (or even evidence for interstellar cooperation) which forms the basis of the so-called Fermi Paradox [8-10], the observation that the Earth has not been colonised by other civilisations despite having being wide open to interference from outside for billions of years. It is too early to tell, but it may be that the Fermi Paradox is telling us that the Galaxy is empty (or largely empty) of intelligent life, and that future human (or post-human) interaction with it may occur along the lines Stapledon sketched for an uninhabited Solar System. We will return to this point in Section 9 below.

**4. If the planets are uninhabited**

If the planets are uninhabited, Stapledon argues that it is still desirable that their exploration and colonisation be undertaken by a united humanity in order to avoid proliferating national rivalries throughout the Solar System. Clearly this is not the way space exploration has actually proceeded to-date, although the recent formulation of the Global Exploration Strategy [11] gives at least some hope that we may be moving in this direction.

As for motives for planetary exploration, Stapledon here for the first time acknowledges scientific curiosity as a motivating factor. However, consistent with his later sidelining of science as a major driving force for space activity (see Section 6), he immediately goes on to state that science is more likely to be the ostensible than the real motivation for planetary exploration. Whereas the history of space exploration to-date might be taken to be consistent with this view, with nationalistic competition often acting as the underlying 'real' motive, this was not what Stapledon had in mind because he had already decided that space exploration should be undertaken by a united humanity. Rather he thought that the underlying motive was altogether a more noble one, namely the inherent human spirit of adventure. As he put it:

> *"Bold young people would be very ready to give their services for planetary exploration. Their effective motive would probably be sheer adventure, though the rational justification of such costly and dangerous undertakings would of course be the advance of science … The irrational, romantic glamour of opening up unexplored worlds will be too strong, even if those worlds turn out to be inhospitable and dreary wastes."*

Stapledon does not make the point explicitly here, but as I have argued elsewhere [6], the 'sheer adventure' of exploring other worlds may fulfil a useful social function in the context of a politically united humanity. This is because, as Cleaver [5] also recognized, it could act as the kind of 'moral equivalent of war' advocated by the philosophers William James [12] and Bertrand Russell [13]. Russell's formulation appears particularly germane in this context, *viz*:

> *"If the world is ever to have peace, it must find ways of combining peace with the possibility of adventures that are not destructive"* [13]

What better source of non-destructive adventure could there be for a united humanity than the exploration and colonisation of other planets?

Stapledon acknowledges that the surface conditions of other planets are not likely to be immediately suitable to support human life, and proposes modifying the environments of the Moon, Mars, and Venus in order to render them habitable. Here he builds on ideas first expressed in '*Last and First Men*' [3], in what was one of the earliest published allusions to what we would now call 'terraforming':

> "*It was necessary either to remake man's nature to suit another planet, or to modify conditions upon another planet to suit man's nature.*" [3; p. 225]

As this quotation makes clear, Stapledon was also open to the possibility of modifying human physiology, as well as modifying planetary environments, to enable planetary colonisation.

**5. Adapting man to the planets**

Indeed, one of the key insights in this part of Stapledon's lecture was the realisation that relying on physical terraforming alone is unlikely to be sufficient to render other planetary environments habitable for human beings. That is to say, while physical terraforming may go part of the way, it will probably also be necessary to adapt human physiology to the newly created environments. As he put it:

> "*If the planets are unadaptable to man in his present form, perhaps man might adapt himself to the alien environments of those strange worlds. Or rather, perhaps a combination of the two processes might enable man to make the best possible use of those worlds. In fact, given sufficient biological knowledge and eugenical technique, it might be possible to breed new human types of men to people the planets.*"

Since Stapledon wrote this our 'biological knowledge' has of course increased enormously, and the relatively crude selective breeding implied by 'eugenical technique' would be superseded by genetic engineering. The ever accelerating pace of our capabilities in this area bring to mind Stapledon's analogy of a river approaching a precipice – we cannot see where it is going, and it is likely to have very significant implications for the future of humanity whether or not we colonise other planets. All one can say is that it appears highly likely that if, at some future date, we choose to genetically modify human beings so as to adapt them to partially terraformed planetary environments then we will have the technical means to do so. Indeed, based on current progress, it seems plausible that we will develop these genetic capabilities *before* we are able to alter the physical environments of other planets.

Stapledon considered terraforming Venus, Mars, and the Moon. Of these, both Venus and Mars were thought to be more Earth-like in 1948 than they are now known to be. Venus, in particular, would require such massive physical terraforming efforts before even genetically adapted humans could live there [14] that human colonisation appears unlikely for millennia, if at all. The Moon would also require massive technological intervention to introduce an atmosphere (presumably brought in from the outer Solar System), and would require continuous intervention to maintain it owing to the low gravity. Even Mars, although it appears to be the most easily terraformable planet in our Solar System, would also probably also take millennia to render habitable [14]. However, at least in this case, one could imagine that the ability to "breed, or otherwise construct, human or quasi-human races adapted to strange environments" might shorten the timescale.

There is one further point to be made -- just as Stapledon's discussion of human interactions with inhabited planets is more likely to be relevant in the context of future galactic colonisation than to our own Solar System, so is his discussion of terraforming and the genetic adaptation of humanity to novel environments. Based largely on results from the *Kepler* mission [15,16], it is becoming clear that planets that are broadly 'Earth-like' in the sense of being rocky planets of roughly an Earth mass within the habitable zone are probably quite common in the Galaxy (see [17] for a recent review). We do not yet know if any of these are inhabited by indigenous life-forms, but the Fermi Paradox [8,9] seems to suggest that such planets only rarely give rise to space-faring civilisations. Many of these planets might be habitable to terrestrial life genetically engineered to match their environments, and it is not beyond the bounds of possibility that, with or without the benefit of physical terraforming, some 'Earth-like' exoplanets might be colonisable by genetically adapted 'quasi-human' species ultimately originating from Earth. Of course, such possibilities lie far in the future, but they are entirely compatible with Stapledon's vision for interplanetary colonisation as enunciated in 1948.

**6. What is it all for?**

Stapledon now comes to what he acknowledges to be "the real crux" of his subject:

> *"Would there be any point in colonizing the planets? ... What is it all for?*
> *Why not just stay put on our native planet and muck along as before?"*

Interestingly, and despite having referred to it in a positive light earlier, he explicitly excludes scientific curiosity as a significant motivation, arguing that "though it might be an important motive for some individuals, is not likely to be the determining factor." It has indeed proved to be the case that science is not the only, or even the main, motivation for space activity, and especially not for large human spaceflight programmes like Apollo or the International Space Station (for both of which geopolitical concerns were and are dominant). Nevertheless, science is *a* major beneficiary of space exploration, and I think that Stapledon was mistaken to downplay it as a motivating factor to the extent that he did in his 1948 lecture.

Instead Stapledon goes on to consider three other possible motives for space colonisation:

- To obtain physical resources from the other planets

- To "leave a mark" on the universe

- To "make the most of man" …to enable "the full expression of the most developed capacities of the human species."

Stapledon is dismissive of the idea of exploiting other worlds for their material resources. He does not deny that such resources may exist, and indeed he appears particularly concerned that by exploiting them mankind might become *too* rich and collapse into decadence. He is explicit on this point, stating that

> *"If the fruit of all the devotion of the British Interplanetary Society is to be merely the debauching of mankind with the riches of other worlds, you had better all stop paying your subscriptions."*

It is possible to agree with this view up to a point, but only up to a point. Humanity as a whole is not currently in any danger of becoming too rich, and there is an argument that if we are to provide a world population that may stabilize at 10 billion by 2070 [18] with an adequate standard of living indefinitely, and without destroying Earth's natural environment in the process, then access to the energy and material resources of the Solar System may be helpful [19]. Moreover if, as Stapledon goes on to argue, the ultimate purpose of space colonisation is to maximise the creative potential of humanity then this will require the utilisation of resources on at least the colonised planets (as Stapledon himself recognizes later in his lecture, where he writes that man "should avail himself of their resources in such ways as to advance the expression of the spirit in the life of mankind").

Stapledon appears ambivalent to the second motive for space colonisation that he identifies, namely the common human impulse to make a mark on the world around us. On the one hand, he finds such impulses to be the hallmark of "uncultured minds", but on the other considers it to be "harmless, even worthy" provided that we "make our mark in inoffensive and if possible actually useful ways." Probably we can all agree with him that merely trying to make a mark on the universe for its own sake is not a sufficient motive for space exploration or for anything else; there has to be some higher motive.

This higher motive is to enable "the full expression of the most developed capacities of the human species." Here at last is a cause worthy of the effort that will be involved in colonising other planets. However, in order to develop this concept, it is necessary to get a firmer understanding of what man is all about.

**7. Fundamental values**

Perhaps the most significant insight that Stapledon expressed in his 1948 lecture was the realisation that:

> *"If one undertakes to discuss what man ought to do with the planets, one must first say what one thinks man ought to do with himself."*

That is to say, we first have to identify the fundamental values to which humanity should aspire, and then examine how the colonisation of the planets may help advance those values. In his earlier lecture 'Mankind at the Crossroads' [20], delivered in France the year before 'Interplanetary Man?', Stapledon had stressed that there had to be some transcendent purpose for human society beyond merely looking after ourselves. As he put it:

> *"For no society can be wholesome unless it is orientated to something more than man, or something in addition to the greatest happiness of the greatest number of existing human individuals."* [20; p.216]

He did not deny that seeking 'the greatest happiness of the greatest number' was an appropriate goal for human society, only that it is not *sufficient*. There has to be some higher, more fundamental, purpose.

Stapledon argues that this greater purpose lies in developing human (and eventually post-human) cultural diversity, intellectual and aesthetic potential, and, especially, what he called 'spirituality'. By 'spirituality' he meant a striving for "sensitive and intelligent awareness of things in the universe (including persons), and of the universe *as a whole*" (Stapledon's emphasis) and for "appropriate and creative action in relation to all this." The key point articulated by Stapledon was that this should be the aspiration of all human development anyway, with or without space colonisation, but that the latter would greatly increase the *scope* for developing human potential and would therefore be consistent with the fundamental values he has identified. As he puts it:

> "*It is in this connection that the planets open up new possibilities. If man can establish in some of those other worlds new and specially adapted human or quasi-human races then those races … should develop new expressions of the spirit at present inconceivable to terrestrial man.*"

**8. A commonwealth of worlds**

Stapledon realised that while much of the cultural richness resulting from planetary colonisation would come from the diversity of the colonised worlds, it would nevertheless be desirable to pool this experience in some way. This led him to advocate the development of 'a commonwealth of worlds'. Thus:

> "*the goal for the solar system would seem to be that it should become an interplanetary community of very diverse worlds each inhabited by its appropriate race of intelligent beings, its characteristic "humanity"….. Through the pooling of this wealth of experience, through this 'commonwealth of worlds' new levels of mental and spiritual development should become possible, levels at present quite inconceivable to man.*"

Stapledon does not explicitly address the political organisation of his 'commonwealth of worlds'. However, it seems to me to be implicit in his reasoning that, just as the political unification of humanity is desirable on Earth, so it would also be desirable on an interplanetary scale, and for essentially the same reasons. While diversity is desirable, conflict is not, and certainly not interplanetary conflict where the continued habitability of whole planets may be at stake. It therefore follows that interplanetary cultural diversity will need to be managed within some kind of appropriate political structure.

Political systems that best combine local autonomy with unity at the highest level, such that diversity may be preserved but conflict between local jurisdictions prevented, are those based on federal principles (e.g., [21-23]). Indeed, among political systems, the federal principle appears uniquely appropriate for Stapledon's 'commonwealth of worlds' because it is naturally expandable from local (sub-planetary) to planetary, and, in principle, to interplanetary scales. The inherent peaceable expandability of democratic federal forms of government was in fact recognized to be a positive advantage by the pioneers of American federalism in the eighteenth century, and Alexander Hamilton [24] pointed out that the only alternatives for government on the largest scales would either be tyranny (to which

Stapledon was rightly opposed, but which Cockell [4] has warned is a real risk in an interplanetary context) or the

> *"splitting ourselves into an infinity of little, jealous, clashing, tumultuous commonwealths, the wretched nurseries of unceasing discord and the miserable objects of universal pity and contempt."* [24; p. 73]

Neither alternative is likely to provide the kind of environment within which human potential can be maximised. On the other hand, a federal 'commonwealth of worlds', within which local (both planetary and sub-planetary) diversity can flourish, but interplanetary conflict can be avoided, would be the kind of political arrangement most consistent with Stapledon's higher aspirations for interplanetary man, at least on the scale of the Solar System. Whether a federation, or any other kind of unified system of government, could be extended to an *interstellar* 'commonwealth of worlds', to which Stapledon turns next, is far more doubtful given the time delays that will be inherent in any form of interstellar communication.

**9. Man and Cosmos**

In the final section of his lecture Stapledon considers humanity in a wider, cosmic context. He notes that at the time he was speaking "interstellar travel seems to us the wildest fantasy" but that

> *"we should not entirely rule out the possibility that a human race far more advanced than ourselves might some day travel far beyond the limits of the solar system."*

Indeed, if his earlier writings are anything to go by, Stapledon actually thought that, given appropriate technology, interstellar travel would be all too possible:

> *"Interstellar, as opposed to interplanetary, travel was quite impossible until the advent of sub-atomic power. …. Immense exploration vessels …. could be projected by rocket action and with cumulative acceleration till their speed was almost half the speed of light. … Races that had attained and secured a communal consciousness would not hesitate to send out a number of such expeditions."* [7; pp. 141-143]

The practicality of interstellar travel raises the possibility both of human (and post-human) colonies expanding out among the stars, and the possibility of contact with extraterrestrial intelligences (ETI). The probability of the latter will clearly depend on how common ETI actually are in the Galaxy. As discussed in Section 3, if ETI are common then the various modes of their possible interaction with humanity are essentially the same as those Stapledon has already considered in the context of inhabited planets in our own Solar System. In his 1948 lecture Stapledon seems to assume that ETI will be common and interaction with humanity likely. This leads him to take the optimistic view that a united humanity will at some point

> *"enter into mutual understanding and appreciation with them, for mutual enrichment and the further expression of the spirit. One can imagine some sort of cosmical community of worlds."*

Stapledon considers such a 'cosmical community', which in many ways prefigures Bracewell's later concept of a 'Galactic Club' [25], to be desirable because it would extend to a vaster scale the trend of increasing 'diversity in community' and opportunities for 'spirituality' (as Stapledon defines it) that he advocates for planetary colonisation within the Solar System. Moreover, in what is perhaps the deepest foray into philosophy in his 1948 lecture, Stapledon argues that it is only through intelligent 'awakened' beings that the inanimate Universe can know itself, and that

> *"the ultimate goal of all awakened beings must inevitably be (how can one least misleadingly put it?) the expression of the objective cosmos in subjective experience and creative action, the fulfilment of the cosmos in cosmical awareness."*

Maximising the spread of subjective awareness of, and creative action within, the objective Universe is important because the Universe's ability to produce thinking beings with these capabilities will decrease with time as the stars run down and entropy increases. There is therefore only a relatively narrow window of opportunity for the Universe to know itself before it is too late. Or, as he puts it, there is a

> *"race between cosmical fulfilment and cosmical death, between the complete awakening of consciousness in the cosmos, and eternal sleep."*

It follows that the more intelligent civilisations there are in the Universe, and the better they are coordinated in a 'cosmical community', the more the universe will be able to know itself before it slides into oblivion. In Stapledon's view this is a transcendent and, in some sense, an *absolute* good. Others, quite possibly influenced by Stapledon, have reached the same conclusion subsequently. Sagan [26] may have put it best:

> *"We are the local embodiment of a Cosmos grown to self-awareness. ... Our obligation to survive is owed not just to ourselves but also to that Cosmos, ancient and vast, from which we spring."* [26; p. 374].

I think that this view of the transcendent importance of rational beings as providing the means by which the Universe knows itself is another of the key insights articulated in Stapledon's lecture. However, while we may agree that the more diversity and intelligence there is the richer the universe will be, and if some of these intelligences are able to form 'cosmical communities' the richer those communities will be, it is not obvious that the linking of these communities is actually required to achieve the highest aim of the universe maximising knowledge of itself. Arguably, what matters in that context is that the *number* of intelligent civilisations be as large as possible, not necessarily that they are linked together in a cosmical community. This may be just as well because, although Stapledon is adamant that all sufficiently evolved 'awakened' beings will share the same set of values, and will therefore be able to benefit from coordinated 'spiritual' activity (he goes so far as to describe any other view as "nonsense"), this is really little more than an assertion on his part. It would seem at least as likely that independently evolved intelligences, having entirely different biologies, never mind ethical systems, would be *too* diverse to enable constructive interaction between them.

However, all this may be moot because, as discussed in Section 3, we have no evidence that the Galaxy actually contains other intelligent civilisations with whom we could join in a 'cosmical community'. On the contrary, the Fermi Paradox may indicate that other civilisations with whom we might interact are actually rare to non-existent [8-10]. Indeed, in *Star Maker* [7; p.140], Stapledon himself pointed out that the probability of intelligence evolving is likely to be an (unknown) function of galactic age and, depending on the (equally unknown) timescales involved, it is therefore entirely possible that few, if any, other intelligent civilisations have arisen in the Galaxy before our own. This would certainly be consistent with the Fermi Paradox and the negative SETI detections to-date [9].

If our Galaxy really is empty of other intelligent (or, more strictly, technological) civilisations then it follows that the future of intelligence in the Galaxy, or at least our part of it, will depend on *us*. The logic of Stapledon's whole argument, and one with which I broadly agree, is that it would then be desirable for humanity (or post-humanity) to start moving out through the Galaxy colonising uninhabited planets because this would enhance the diversity and creative potential of life in the Galaxy and, as an unintended but inevitable by-product, also increase the opportunities for the Universe to know itself. We must however recognize, as Stapledon foresaw, that these other planetary environments are unlikely to be such as to support human life directly, and that terraforming and/or genetic engineering of the colonists is likely to be necessary.

Even if the post-human colonists soon cease to look or feel much like us, the fact that they will all ultimately have had a common origin, and therefore at least a common underlying biology (and perhaps also certain common ethical perspectives), may make forming a 'cosmical community' between them much easier than would be the case between independently evolved intelligences. The extent to which this would be practical given the distances involved and the finite speed of light remains to be seen, and would depend on (i) whether or not the speed of light indeed turns out to be an absolute limit to the maximum speed of communication (which, contrary to popular belief, is still not actually known with confidence [27]); and (ii) the longevity of the participating cultures. Even if the speed of light is an absolute limit (which is the safest assumption given current knowledge), if the participating cultures have lifetimes of hundreds of thousands or millions of years then interstellar 'cosmical communities' may still be possible, as Stapledon himself recognized in *Star Maker*:

> *"Even at such a rate of travel* [0.5c] *voyages to the nearer stars were well worth undertaking … It must be remembered that a fully awakened world had no need to think in terms of such short duration as a human lifetime. Though its individuals might die, the minded world was in a very important sense immortal. It was accustomed to lay its plans to cover periods of many million years."* [7; pp. 142-3]

Given such longevity, 'cosmical communities' of the kind envisaged by Stapledon would be possible on at least a galactic scale, and might be possible even with a cluster of galaxies such as the Local Group. It seems, however, that cosmical communities organised on a larger scale than this would be unlikely unless faster-than-light travel and/or communication proves to be possible [27].

## 10. Conclusions

Having reviewed Stapledon's lecture on "Interplanetary Man?" what can we take from it today? In my view, the lecture contains a number of key insights which are just as valid now as they were in 1948. These are:

(1) That "if one undertakes to discuss what man ought to do with the planets, one must first say what one thinks man ought to do with himself";

(2) That this ultimate objective for humanity involves maximising human unity, well-being, cultural diversity, and intellectual and aesthetic potential;

(3) That these objectives are desirable anyway, with or without space exploration, but that space exploration would greatly increase the *scope* for developing human (and later post-human) potential;

(4) That the same logic which calls for uniting humanity on Earth applies equally beyond it, hence the desirability of a 'Commonwealth of Worlds'; and

(5) That in some deep sense the expansion of intelligence into the Universe may have wider *cosmic* significance by increasing the extent to which the Universe becomes aware of itself.

On the other hand, there are also some aspects of the lecture that I think we could take issue with. For example, Stapledon tended to down-play economic and scientific motivations for space exploration, yet the former is important for maximising human well-being and the latter is a key component of human intellectual development. Stapledon also tended to view the unification of humanity, and the improvement of the human condition, as prerequisites for space exploration and colonisation. However, while desirable, these are probably better viewed as *aspirations* which can (and should) be pursued in parallel.

That said, for reasons given elsewhere [6], I think Stapledon was right to point to synergies between human political unification and space exploration, and I think he would have been pleased to see recent attempts to internationalise space activity through the Global Exploration Strategy [11] and the resulting Global Exploration Roadmap [28]. These are a long way from creating either a united humanity or a 'Commonwealth of Worlds', but they are at least a major step towards the goal of a cooperative, global, space exploration programme that would be entirely congruent with Stapledon's overarching vision as set out in "Interplanetary Man?"